\documentclass[aps,reprint,superscriptaddress,
nofootinbib,showpacs,amsmath,amssymb]{revtex4-1}

\usepackage{latexsym,graphicx,slashed,bm,dcolumn}

\newcommand{\be}[1]{\begin{equation}\label{#1}}
\newcommand{\beq}{\begin{equation}}
\newcommand{\eeq}{\end{equation}}
\newcommand{\beqn}[1]{\begin{eqnarray}\label{#1}}
\newcommand{\eeqn}{\end{eqnarray}}

\newcommand{\dub}[2]{\left(\begin{array}{c}{#1}\\{#2}
\end{array}\right)}

\renewcommand{\to}{\rightarrow}

\def\ov{\overline}
\def\ee{\end{equation}}

\def\lsim{\raise0.3ex\hbox{$\;<$\kern-0.75em\raise-1.1ex
\hbox{$\sim\;$}}}
\def\gsim{\raise0.3ex\hbox{$\;>$\kern-0.75em\raise-1.1ex
\hbox{$\sim\;$}}}
\def\cal{\mathcal}

\def\cL{{\cal L}}
\def\cM{{\cal M}}
\def\cN{{\cal N}}
\def\cO{{\cal O}}

\def\tphi{\bar{\phi}}

\def\tf{\bar{f}}

\def\tq{\bar{q}}
\def\tl{\bar{l}}

\def\tu{\bar{u}}
\def\td{\bar{d}}

\def\lpr{l^\prime}
\def\phpr{\phi^\prime}

\def\rB{{\rm B}}
\def\rL{{\rm L}}

\begin{document}

\title{ Neutron lifetime  and dark decays of the neutron and hydrogen    }

\author{Zurab~Berezhiani}
\email{E-mail: zurab.berezhiani@lngs.infn.it}
\affiliation{Dipartimento di Fisica e Chimica, Universit\`a di L'Aquila, 67100 Coppito, L'Aquila, Italy} 
\affiliation{INFN, Laboratori Nazionali del Gran Sasso, 67010 Assergi,  L'Aquila, Italy}


\begin{abstract} 
The neutron, besides its $\beta$-decay $n\to p e\bar\nu_e$,  might have a new decay channel 
$n\to n' X$ into mirror neutron $n'$, its nearly mass degenerate twin from  parallel dark sector, 
and a massless boson $X$ which can be  ordinary and mirror photons  or some more exotic particle.  
Such an invisible decay could alleviate the tension between the  neutron lifetimes measured  
in the beam and trap experiments. 
I discuss some phenomenological and astrophysical consequences of this scenario, 
which depends on the mass range of mirror neutron $n'$.  
Namely, the case  $m_{n'} < m_p + m_e$ leads to a striking possibility is that the hydrogen atom 
$^1$H (protium), constituting 75 per cent of the baryon mass in the Universe,  could in fact be unstable:  
it  can decay  via the electron capture into $n'$ and $\nu_e$, 
with  relatively short lifetime $\sim 10^{21}$ yr or so. 
If instead $m_{n'} > m_p + m_e$, then the decay $n'\to pe\bar \nu_e$ is allowed and $n'$ can 
represent an unstable dark matter component 
with rather large  lifetime exceeding the age of the Universe. 
Nevertheless, this decay would produce substantial diffuse gamma background.     
The dark decay explanation of the lifetime puzzle, however, 
has a tension with the last experimental results measuring $\beta$-asymmetry in the neutron decay. 
\end{abstract}

\maketitle

\noindent
{\bf 1.} 
The neutron, a long-known particle which constitutes half of the mass in our bodies, 
may still reserve many surprises. While the free neutron is unstable,
 there still remains a problem  to understand what is its true lifetime. 
According to to Standard Model  (and common wisdom of baryon conservation) 
the neutron can have only  $\beta$-decay $n\to pe\bar\nu_e$ (including the subdominant 
daughter branch of radiative decay $n\to pe\bar\nu_e \gamma$ with the photon emission). 
Hence, its lifetime can be measured in two different ways, known as the trap and beam methods. 
The trap experiments are  based on disappearance of the ultra-cold neutrons (UCN) stored  
in material or magnetic traps.  They measure the true lifetime $\tau_n$, 
equivalent to its total decay width  $\Gamma_n  = \tau_n^{-1}$,  via counting the survived UCN  
for different storage times and reproducing the exponential time dependence $\exp(-t/\tau_n )$ 
after accurately estimating and subtracting other effects of the UCN losses related
 to the wall absorptions, up-scattering etc. 
The beam experiments are the appearance experiments, measuring the $\beta$-decay width 
$\Gamma_\beta$ by counting the protons produced via decay $n\to pe\bar\nu_e$ 
in the monitored beam of cold neutrons. 
Clearly, in the absence of new physics  both methods should measure the same value,    
$\Gamma_n = \Gamma_\beta$. 

However, as it was noticed quite a time ago \cite{Serebrov:2011}, 
the neutron lifetimes measured with two methods have some discrepancy.  
Careful re-analysis of the previous experimental results and new 
measurements with increased precision rendered this discrepancy more evident \cite{Wietfeldt:2014}. 

\begin{figure}
\includegraphics[width=0.50\textwidth]{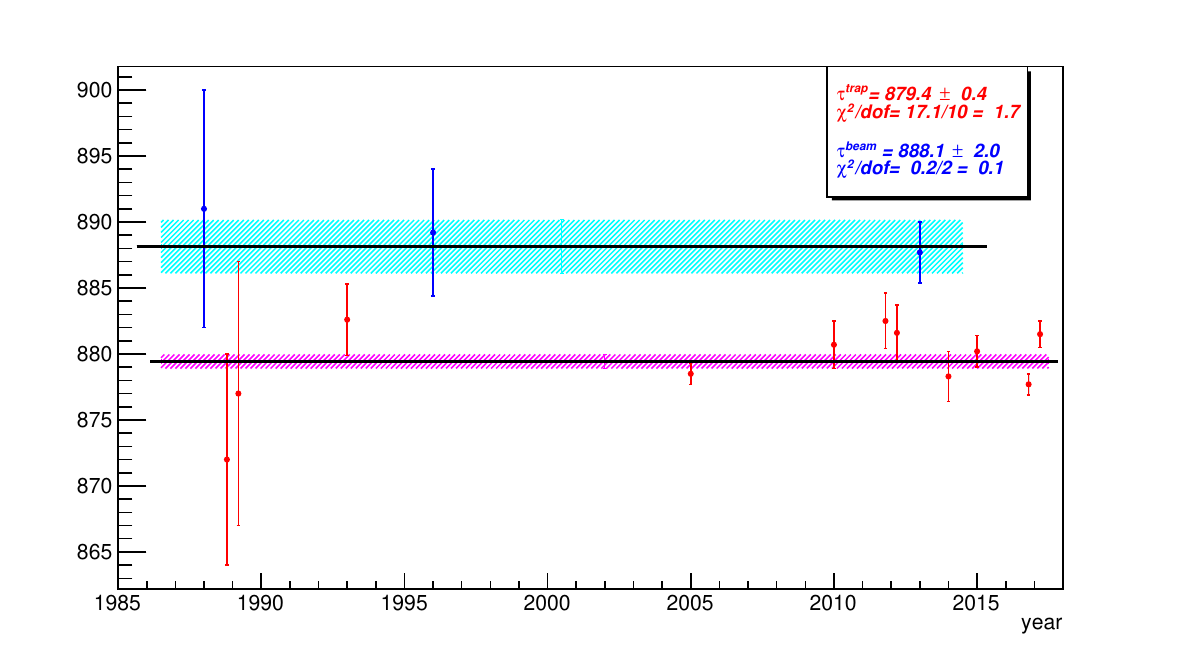}
\caption{Results of the trap (red) and beam (blue) measurements and respective averages. }
\label{fig:tau}
\end{figure}

Fig. \ref{fig:tau} summarizes  results of the neutron lifetime measurements performed 
from 1988 till now 
(experiments which results were removed and the ones reporting error-bars exceeding 10~s are not included). 
The trap experiments, Refs. 
\cite{Kharitonov,Paul,Mampe:1993,Serebrov:2005,Pichlmaier:2010,Steyerl:2012,Arzumanov:2012,
Ezhov:2014,Arzumanov:2015,Pattie:2017,Serebrov:2017}, 
are in good agreement with each other, and their average reads
\be{trap} 
\tau_{\rm trap} = (879.4 \pm 0.5)~{\rm s} .
\ee 
The results of beam experiments, Refs. \cite{Spivak:1988,Byrne:1996,Yue:2013}, also 
are in fine agreement and their average yields 
\be{beam} 
\tau_{\rm beam} = (888.1 \pm 2.0)~{\rm s} .
\ee 
The discrepancy is about 9 s. Formally the beam result (\ref{beam}) is $4\sigma$ 
away from the trap result (\ref{trap}). 

It is instructive to follow the time evolution of the neutron lifetime as it is reflected 
in the Particle Data Group (PDG) editions of last years. 
PDG 2010 \cite{PDG2010}  summarises  available experimental results 
and adopts the world average $\tau = 885.7 \pm 0.8$~s.  
However, it discards the  result  of most accurate measurements  
 reported by the Serebrov's group in 2005, $\tau = 878.5 \pm 0.8$~s  \cite{Serebrov:2005},   
with the following comment:  
{\it ``SEREBROV 05 result is $6.5\sigma$ deviations from our average of previous results 
and $5.6\sigma$ deviations from the previous most precise result (that of ARZUMANOV 00)''},  
since by that time results of all  beam and trap experiments (excluding that of Serebrov's)  
were in good agreement.
However, already the next edition, PDG 2012  \cite{PDG2012},
 adopted a world average $\tau = 880.1 \pm 1.1$~s, rather distant 
from the previous PDG 2010 one, and also with larger error-bars.   
This value suffered only minor changes in following PDG editions \cite{PDG2014,PDG2016}. 
Namely, PDG 2018 quotes its value as $\tau = 880.2 \pm 1.0$~s \cite{PDG2018}, 
without the latest results of Refs. \cite{Ezhov:2014,Pattie:2017,Serebrov:2017} being included. 
What has happened between the PDG 2010 and PDG 2012 editions? 
First, the result of Serebrov's experiment  \cite{Serebrov:2005} was included, 
and second, in 2010 
Serebrov and Fomin critically reanalyzed the results of all trap experiments performed before 
2005  and found a systematic error of about 6~s \cite{Serebrov:2010}. 
In consequence, many experimental groups themselves reevaluated their previous results and adopted  
new corrected values (see Refs. \cite{Steyerl:2012,Arzumanov:2012}). 
By the time, also results of new trap measurements  \cite{Pichlmaier:2010} were published, 
all consistent with the previously discarded result of Ref. \cite{Serebrov:2005}. 

On the other side, the beam results showed quite an opposite trend.  The re-analysis of 
previous beam measurements brought to larger value of $\tau_\beta$ 
with smaller error-bars \cite{Yue:2013}.  
In this way, the discrepancy between the neutron lifetimes measured in the trap and beam experiments 
became rather evident what renders the situation more enigmatic. 
 
 \medskip 
 \noindent{\bf 2.} 
 The fact that $\tau_{\rm beam}$ (\ref{beam})  is larger  than  $\tau_{\rm trap}$ (\ref{trap}) 
 with about one per cent difference:  
 \be{Delta-tau}
 \Delta\tau_n = \tau_{\rm beam}  - \tau_{\rm trap} = (8.7 \pm 2.1)~{\rm s} , 
 \ee 
 may suggest that apart of usual $\beta$-decay $n\to pe\bar\nu_e$, 
the neutron may have a new decay channel, invisible or  semi-invisible 
(i.e. in principle detectable but not yet excluded experimentally).
%
In this case, the trap experiments would measure the neutron total decay width, 
$\Gamma_n = \Gamma_\beta + \Gamma_{\rm new} = \tau_{\rm trap}^{-1} = 7.485 \times 10^{-28}$~GeV,  
where $\beta$-decay width $\Gamma_\beta = \tau_{\rm beam}^{-1}$ measured by beam experimentts 
should constitute a dominant part of it, with the branching ratio  
${\rm Br}(n\to pe\bar\nu_e) = \Gamma_\beta/\Gamma_n = \tau_{\rm trap}/\tau_{\rm beam} = 0.99$. 
Therefore, the new decay channel with about 1 per cent branching ratio    
would suffice for resolving the discrepancy. Namely, given the error-bars in (\ref{trap}) and (\ref{beam}), 
for reconciling the situation at about $1\sigma$ level 
one would need   
\be{Gamma-new}
\Gamma_{\rm new} = (7 \pm 2) \times 10^{-30}~{\rm GeV}.
\ee
E.g., the neutron could decay in two invisible particles, $n\to n'+X$, where $n'$ is a ``dark" fermion 
with a mass $m'_n < m_n$  and $X$ is a massless (or light enough) ``dark" boson, 
while the ``yet-invisible'' mode could be $n\to n'+ \gamma$ with the photon emission.

Clearly, new particle $n'$ cannot be arbitrarily light, and the mass splitting $\Delta m = m_n-m_{n'}$ is 
is limited  by the stability of chemical elements  with precision of about a MeV. 
While allowing the decay $n\to n'+X$ for a free neutron, i.e. $m_{n'} < m_n$, 
this decay should be forbidden for a neutron 
bound in that nuclei which are known to be stable. 
The strongest bound comes from the stability  of  $^9$Be which has a rather fuzzy nuclei, 
having the minimal neutron separation energy among all stable elements.   
Transition $n\to n'$, if allowed by phase space, would transform $^9$Be, $M(^9{\rm Be}) = 8394.79535$~MeV,
 into $^8$Be, $M(^8{\rm Be}) = 7456.89447$~MeV, 
which is $\alpha$-unstable  with decay time $\sim 10^{-16}$~s. 
In fact, the stability of $^9$Be atom against the removal of extra neutron $n \to n'$  
implies that $m_{n'}$ should be larger than the mass of $^9$Be
minus twice the mass of $^4$He, $M(^4{\rm He}) = 3728.40132$~MeV. 
In this way, one can set a lower limit 
\be{Be}
 m_{n'} >  937.992~{\rm MeV}, \quad {\rm i.e.} \quad \Delta m_{\rm max}  = 1.573~{\rm MeV}
\ee
Other stable  elements do not give competitive limits. E.g. Deutrium $^2$H stability implies 
$m_{n'} > M(^2{\rm H}) - M(^1{\rm H}) = 937.3358$~MeV or $\Delta m < 2.230$ MeV,  
while the limits from other elements are yet weaker. 

A sterile particle $n'$ so closely degenerate in mass with the neutron, with precision of 
$\Delta m/m_n\sim 10^{-3}$ can be introduced 
{\it ad hoc} as an elementary fermion but this sort of fine tuning does not look very appealing. 
In addition,  for $n'$ being an elementary fermion with 
negligible self-interaction, rapid $n\to n'+X$ transition would have disastrous consequences 
for the neutron star stability:   
the stars made of degenerate gas of free fermions can have a maximal 
mass $M_{\rm max} = 0.627~M_\odot (1~{\rm GeV}/m_{n'})^2$ \cite{Narain}  
which for $m_n'\simeq m_n$ gives $M_{\rm max} \approx 0.71~M_\odot$.\footnote{In fact, this
is well-known since 1939 from the original work of Oppenheimer and Volkoff  \cite{OV}   
which  obtained $M_{\rm max} \simeq 0.7~M_\odot$ for maximal mass neutron stars
 considering neutrons as degenerate Fermi gas 
(see also in textbook of Shapiro and Teukolsky \cite{Shapiro}).} 
But the typical masses of neutron  stars are about twice as larger, and moreover 
at least two pulsars of  $2\,M_\odot$ were observed. For achieving that large masses, 
one has to introduce strong repulsive  force between fermions $n'$ mediated by  
a vector boson $I$ with large coupling constant $g$. 
In this case the maximal mass can be increased to 
$M_{\rm max} = (0.7 +0.3 y_{n'})(m_n/m_{n'})^2 ~M_\odot $  
where the parameter $y= (g/\sqrt2) m_{n'}/m_I $ describes the interaction strength \cite{Narain}. 
In the case for the neutrons this repulsive force can be induced by $\rho-\omega$ mesons,  
and taking $m_\rho \simeq 0.8$~GeV and $g_{\rho n} \simeq 13$, one has $y_n\simeq 10$. 
In this case the compact objects can be as heavy as $3.3\,M_\odot$.\footnote{Of course, 
for real neutron stars this is only a rough approximation. 
In a realistic approach, along with the repulsive vector interaction also attractive scalar interaction
should be included  which reduces $M_{\rm max}$. 
Unfortunately, the true equation of state of dense 
nuclear matter is of difficult determination and real value of maximal mass of neutron stars 
remains unknown, though experimentally we know that it should be larger than $2\,M_\odot$. }
This suggests that for avoiding the neutron star destabilization due to rapid  $n\to n'X$ decay, 
$n'$ can  be considered as composite fermion subject to short-distance repulsive forces 
of nearly the same strength as the repulsive forces acting between the neutrons. 
Namely, one can envisage that $n'$ is composed of three hypothetical `quarks' 
bounded by some new color $SU(3)$ forces. However, there remains a question 
why it is so nearly degenerate in mass with the neutron.  

In this view, it would natural to consider  new fermion $n'$ as a 
dark twin of the neutron $n$ from a  parallel hidden sector, coined as mirror world, 
 which is an identical copy of ordinary particle sector (for reviews, see e.g. Refs. \cite{Alice,Foot}). 
 The identical form of two sectors can be ensured by a discrete symmetry, $Z_2$ parity.  
In this picture all ordinary particles: the electron $e$, proton $p$, neutrinos $\nu$ etc., 
should have exactly mass-degenerate invisible twins: $e'$, $p'$, $\nu'$, etc. 
which are sterile to our strong and electroweak interactions $SU(3)\times SU(2)\times U(1)$ 
but have their own gauge interactions $SU(3)'\times SU(2)'\times U(1)'$. 
Mirror matter, with its features of being baryon-asymmetric, atomic and thus dissipative, 
can represent part or even entire amount of dark matter in the Universe, 
with specific implications  for the cosmological evolution, 
formation and structure of galaxies and stars, etc. \cite{BDM,BCV,BCV2}. 

Interestingly, the baryon asymmetries in both ordinary and mirror worlds can be
generated by particle processes that violate $B\!-\!L$ and $CP$ in both sectors ~\cite{BB-PRL} 
which can explain the relation between the dark and visible matter fractions in the
Universe, $\Omega'_B/\Omega_B \simeq 5$. 
On the other hand, the same interactions can induce mixing phenomena between ordinary and mirror
particles.  In fact, any {\it neutral} particle, {\it elementary or composite}, may have a mixing with its 
mirror twin. E.g.\ three ordinary neutrinos $\nu_{e,\mu,\tau}$ 
can be mixed with their mirror partners $\nu'_{e,\mu,\tau}$ which in fact are most natural candidates 
for the role of sterile neutrinos \cite{ABS}. 

The mixing between the neutron $n$ and its mirror twin $n'$ was introduced in Ref. \cite{BB-nn'},   
assuming that two states are exactly degenerate in mass, and   
its astrophysical and cosmological implications were discussed in Refs. \cite{BB-nn',nn'-cosm}.  
This mixing is similar, and perhaps complementary, 
 to neutron-antineutron ($n-\bar n$) mixing \cite{Phillips}.  
 However,  in difference from the latter, it is not restricted by the nuclear stability limits 
 since $n-n'$ transition  for a neutron bound in nuclei  cannot take place simply because of 
 energy conservation \cite{BB-nn'}. 
Possible experimental strategies for searching $n-n'$ oscillation were discussed in 
Refs. \cite{BB-PRL,nn'-exp}, and the results of several dedicated experiments 
can be found in Refs. \cite{Experiments}. 

In this paper we consider the situation when mirror symmetry is softly or spontaneously 
broken and $n$ and $n'$ states are not exactly degenerate in mass but have a mass splitting 
of about a MeV, so that $n\to n' X$ decay of free neutron becomes possible.  
This concept suggests intriguing connection  between the neutron lifetime  and dark matter puzzles. 
In particular, dark matter can present entirely in the form of mirror neutrons, without an atomic component, 
 if mirror proton $p'$ is heavier than mirror neutron $n'$,  and it decays as $p' \to n' \bar e' \nu_e'$.  
We shall discuss implications of $n\to n'$ decay in the light of the neutron lifetime puzzle 
and other issues as are the matter stability, dark matter decay, etc. 

\medskip 

\noindent
{\bf 3.}  
One can consider a theory based on the product $G\times G'$ of two 
identical gauge factors (Standard Model 
$SU(3)\times SU(2)\times U(1)$ 
or some its extension),   
ordinary  (O) particles belonging to $G$ and mirror (M) particles to $G'$ (mirror Standard Model 
$SU(3)'\times SU(2)'\times U(1)'$ 
or its equivalent extension). 

In the Standard Model 
the quark fields are represented as Weyl spinors,  the left-handed (LH) ones  
 transforming as weak isodoublets 
and the right-handed (RH) ones as isosinglets, 
whereas the anti-quark fields $\bar q$ which are CP conjugated to $q$  
($\tq_{R,L} = C \gamma_0 q_{L,R}^\ast$)   
have  the opposite chiralities and opposite gauge charges:
\beqn{SM-L} 
& q_L = \dub{u_L}{d_L}, 
~~~
q_R \, = \, u_R, ~ d_R,    \nonumber \\
& \tq_R = \dub{\tu_R}{\td_R}, ~ 
~~~
\tq_L \, = \, \tu_L, ~ \td_L 
\eeqn
(the family indices are suppressed, and the lepton fields are omitted for brevity). 
In addition, we assign to quarks $q_L,u_R,d_R$ a global baryon charge $\rB =1/3$. 
Then antiquarks $\tq_R,\tu_L,\td_L$ have $\rB=-1/3$.  

The parallel M sector $G^\prime =SU(3)^\prime\times SU(2)^\prime\times U(1)^\prime$  
has the analogous quark content  
\beqn{SM-Lpr} 
& 
q^\prime_L = \dub{u'_L}{d'_L} , ~ 
~~ 
q'_R \, = \, u'_R, ~d'_R   \nonumber \\
& 
\tq'_R = \dub{\tu'_R}{\td'_R}, ~ 
~~ 
\tq'_L \, = \, \tu'_L,~\td'_L 
\eeqn
For definiteness, we name  $\tq'_R,\tu'_L,\td'_L$ as mirror quarks and assign 
them a mirror baryon number  $\rB'=1/3$. 
Then  mirror anti-quarks  $q'_L,u'_R,d'_R$ have $\rB'=-1/3$.
 
The Lagrangian of two systems  has a generic form  
\be{Lagr}
{\cal L}_{\rm tot} = {\cal L} + {\cal L}' + {\cal L}_{\rm mix}  \, . 
\ee
where ${\cal L}$ and ${\cal L}'$ respectively are the Standard Lagrangians of O and M sectors, 
including the gauge,  Yukawa and Higgs parts, while 
${\cal L}_{\rm mix}$ stands for possible interactions between the particles of two sectors.
The identical forms of ${\cal L}$ and ${\cal L}'$ can be ensured 
by discrete $Z_2$ symmetry under the exchange $G\leftrightarrow G'$
when all O particles (fermions, Higgs and gauge fields) exchange places with their 
M twins (`primed' fermions, Higgs and gauge fields). 
Such a discrete symmetry can be imposed  {\it with} or 
{\it without} chirality change between the O and M fermions \cite{Alice}.   
However this difference will have no relevance for our further discussion; what is important that 
this symmetry ensures that the gauge and Yukawa coupling constants are the same 
in two sectors.  
Hence, if $Z_2$ symmetry between two sectors is unbroken,  i.e. O and M Higgses $\phi$ and $\phi'$ 
have exactly the same vacuum expectation values (VEVs), 
then mirror world will be an exact replica 
of ordinary particle sector, and all O particles: the electron $e$, proton $p$, neutron $n$ etc., 
would be exactly mass-degenerate with their M twins: $e'$, $p'$, $n'$, etc. 

However, one can envisage a situation when $Z_2$ is spontaneously broken. E.g. one can introduce  
a real scalar field $\eta$ which is odd under $Z_2$ symmetry, i.e. transforms as $\eta \to -\eta$ \cite{BDM}.  
If this scalar acquires a non-zero VEV, then  Its coupling to O and M Higgses 
will give different contributions to their mass terms.  
In this way, the O and M Higgses can get different VEVs, 
and so the masses of O and M quarks would be different. 

Let us consider a situation when each of the O and M sectors 
 is represented by the models with two Higgs doublets 
 $\phi_{1,2}$ and  $\phi'_{1,2}$, responsible for the masses of up and down quarks, as motivated 
 by e.g. supersymmetry.  
 In this case the couplings of $Z_2$-odd scalar 
 $\lambda_1 \eta (\phi_1^\dagger\phi_1 - \phi^{\prime\dagger}_1\phi'_1)$ and  
 $\lambda_2 \eta (\phi_2^\dagger\phi_2 - \phi^{\prime\dagger}_2\phi'_2)$, 
 with $\lambda_{1,2}$ being dimensional couplings,  would lead to different 
 VEVs, $v'_1\neq v_1$ and $v'_2\neq v_2$.   For evading the strong hierarchy problem 
 and related fine tunings, we can assume that all of these four values are in the range of 
 few hundred GeV, with $v_{\rm ew} = (v_1^2+v_2^2)^{1/2}$ determining the (known) 
 ordinary weak scale and 
 $v'_{\rm ew} = (v_1^{\prime 2}+v_2^{\prime2} )^{1/2} \neq v_{\rm ew}$
 determining the mirror weak scale; one can take e.g.  $\lambda_{1,2} \sim 1$~GeV
and  $\langle \eta\rangle$ in the range of few TeV.\footnote{More generically, 
if two sectors contain scalar fields other than the Higgs doublets $\phi_{1,2}$ and $\phi'_{1,2}$, 
as e.g. color scalars $S$ and $S'$ discussed 
in next section,  then the couplings of $Z_2$ odd scalar $\lambda \eta (S^\dagger S - S^{\prime\dagger}S')$ 
would induce different masses for them, $M'_S\neq M_S$. In particular, 
for achieving e.g. $M_S \sim 1$~TeV and $M_{S'} \sim 100$~GeV, we would need rather 
large values of $\lambda$ approaching TeV scale.}
Due to renormalisation group effects, the difference between the O and M Higgs VEVs 
can induce some difference between the QCD scales in two sectors can become somewhat different, 
but for $v'_{\rm ew} \sim v_{\rm ew}$ we expect  that 
$\Lambda'_{\rm QCD} \simeq \Lambda_{\rm QCD}$.  In this case  
the light quark masses of both sectors are expected to be of few MeV and 
so the mass splitting  between M and O nucleons can be in the MeV range. 

In particular, one can envisage a situation when 
$m_{u'} > m_u$ but $m_{d'} < m_d$ and $m_{e'} < m_e$.
 Let us take e.g. a simple example when $v'_1 \simeq 2 v_1$ but $v'_2 \simeq v_2/2$. 
Given that the Yukawa coupling constants in two sectors are the same, 
for the reference masses or our light quarks $m_u\simeq 2$~MeV and $m_d\simeq 4$ MeV, 
the mirror light quarks masses are just inverted,  $m'_u\simeq 4$~MeV and $m'_d\simeq 2$ MeV, 
while for the electrons we have $m'_e \simeq m_e/2$. 
Therefore, it could occur pretty naturally that the mirror neutron and proton have masses 
different from their ordinary twins by a MeV or so, but to different sides arranged as  
$m'_p > m_n > m'_n > m_p$. 
 Namely,  if $m_{p'} > m_{n'} + m_{e'}$, then free M proton $p'$ would be unstable so that mirror world 
 would contain no hydrogen. However mirror neutron $n'$ would be stable and  thus represent 
a self-scattering dark matter with just a perfect cross-section over mass ratio, 
 $\sigma/m_n \sim 1$ bn/GeV.  In the following we consider this case as a reference model. 
 In other possible situation when $\vert m_{p'} - m_{n'} \vert < m_{e'}$, both $p'$ and $n'$ will be stable and  
 one would have dark matter  in two forms, self scattering component $n'$ and dissipative components  
in the form of  mirror hydrogen H$'$ and helium He$'$.

 \medskip 
 
 \noindent {\bf 4}. Let us concentrate on the system of two neutrons, ordinary $n$ and mirror $n'$. 
 The relevant terms of the generic Lagrangian (\ref{Lagr}) are the low energy effective terms 
 related to their masses and magnetic moments: 
\beqn{Lagrangian} 
&& {\cal L} = 
m_n \ov{n} n +  
\frac{\mu_n}{2} F_{\mu\nu} \ov{n}\sigma^{\mu\nu} n   \nonumber \\
&& {\cal L}' = 
m_{n'} \ov{n'} n'  + 
\frac{\mu_{n'}}{2}  F'_{\mu\nu} \ov{n'}\sigma^{\mu\nu} n'  \nonumber \\ 
&& {\cal L}_{\rm mix} = \epsilon \, \ov{n} n' 
+ \frac{\kappa_{nn'} }{2} F_{\mu\nu} \ov{n}\sigma^{\mu\nu} n' 
+   \frac{\kappa'_{nn'} }{2} F'_{\mu\nu} \ov{n}\sigma^{\mu\nu} n' + {\rm h.c.}  
\eeqn 
where $ F_{\mu\nu} = \partial_\mu A_\nu - \partial_\nu A_\mu$ is the electromagnetic 
field strength tensor, and $ F'_{\mu\nu}=\partial_\mu A'_\nu - \partial_\nu A'_\mu$ is the same 
for mirror electromagnetic field, 
$m_n=939.5654$~MeV and $\mu_n=-1.912\mu_N$ respectively are the neutron mass and 
magnetic moment, $\mu_N = e/2m_p$ being the nuclear magneton, and $m_{n'}$ and $\mu_{n'}$  
are those of mirror neutron. We assume that due to $Z_2$ breaking there is a small 
mass splitting between $n$ and $n'$ states, $\Delta m = m_n - m_{n'}\simeq 1$~MeV.  
The magnetic moments $\mu_n$ and $\mu_{n'}$ should also have some tiny difference 
but this is irrelevant for our discussion and one can safely take $\mu_{n'} = \mu_n$ as a good approximation. 
In addition, unlike the case of neutron-antineutron system, transitional magnetic (or electric dipole) moments 
$\kappa_{nn'}$ and $\kappa'_{nn'}$ between the neutron and mirror neutron 
are not forbidden by fundamental symmetry reasons \cite{Arkady}. 
Lagrangians $\cL$ and $\cL'$ in (\ref{Lagrangian}) conserve baryon numbers $\rB$ and 
$\rB'$ separately,  while the mass term in $\cL_{\rm mix}$ mixing the states $n$ ($B=1$) and 
$n'$ ($B'=1$)  conserves the combination of baryon charges  $\ov{\rB}= \rB+\rB'$.

The mixing term ${\cal L}_{\rm mix}$ 
can be induced by the effective six-quark operators with different Lorentz structures 
involving LH and RH quarks 
$u_{L,R}, d_{L,R}$ and mirror antiquarks $u'_{L,R}, d'_{L,R}$ in gauge singlet combinations \cite{BB-nn'}: 
\be{ops}
\frac{1}{\cM^5}(\tu \td \td)(u' d' d')  + {\rm h.c.} 
\ee
The Lorentz, gauge and family indices are suppressed.  
These operators transform the neutron state $n$  (three valent  quarks $udd$, $\bar\rB=1$) 
into mirror neutron $n'$ (three mirror quarks $\tu'\td'\td'$, again $\bar\rB=1$). 
Taking the matrix elements 
$\langle n \vert  udd \vert 0\rangle = K \Lambda_{\rm QCD}^3\simeq K \times 0.015$~GeV$^3$, 
with $K$ being an order 1 coefficient, and equivalently for the mirror neutron, 
we obtain  the $n-n'$ mixing mass as 
$\epsilon \simeq (K/2)^2 (10^{10} \, {\rm GeV}^5/\cM^5) \times 10^{-10}$ MeV. 


Operators (\ref{ops}) can be induced e.g. via seesaw-like mechanism \cite{BB-nn',BM} 
from the following Lagrangian terms:\footnote{Color indices  
and charge conjugation matrix $C$ are suppressed. 
 $u_Ld_L$  can enter in weak isosinglet combination 
$\epsilon^{\alpha\beta}q_{L\alpha}q_{L\beta}$ where $\alpha,\beta=1,2$ are the weak $SU(2)$ 
indices.  For simplicity we take the constants of couplings  $Su_Rd_R$  and $Su_Ld_L$ equal, 
$g_L=g_R=g$, and take $K\simeq 2$. 
}
\beqn{S}
&& \cL = g S u_{R,L} d_{R,L}  +  h_a S^\dagger d_R N_{Ra}  +   {\rm h.c.}   
\nonumber \\
&& \cL' = g S' u'_{R,L} d'_{R,L}  + h_a S^{\prime\dagger} d'_R N_{Ra}'    +   {\rm h.c.} 
\nonumber \\
&& \cL_{\rm mix} = M^{(a)}_{D} N_{Ra} N'_{Ra}  +   {\rm h.c.}; \quad \quad a=1,2,... 
\eeqn 
involving a color-triplet scalar $S$  ($\ov{\rB}=-2/3$) with mass $M_S$  and 
and its mirror partner $S'$ ($\ov{\rB}=2/3$) with mass $M_{S'}$. 
It also involves gauge singlet RH fermions $N_{Ra} $ with 
$\ov{\rB}=-1$ and $N'_{Ra}$  with $\ov{\rB}=1$, 
so that the mass terms in (\ref{S}) conserves the combined charge $\ov{\rB}=\rB+\rB'$. 
In fact, these mass terms  $M_D$ are the Dirac mass terms:  
one can say that RH components $N_R$ of Dirac spinors $N_L+N_R$
 belong to  
ordinary sector and LH components $N_L = C\ov{N'_R}^T$ belong to mirror sector. 

\begin{figure}
\includegraphics[width=0.45\textwidth]{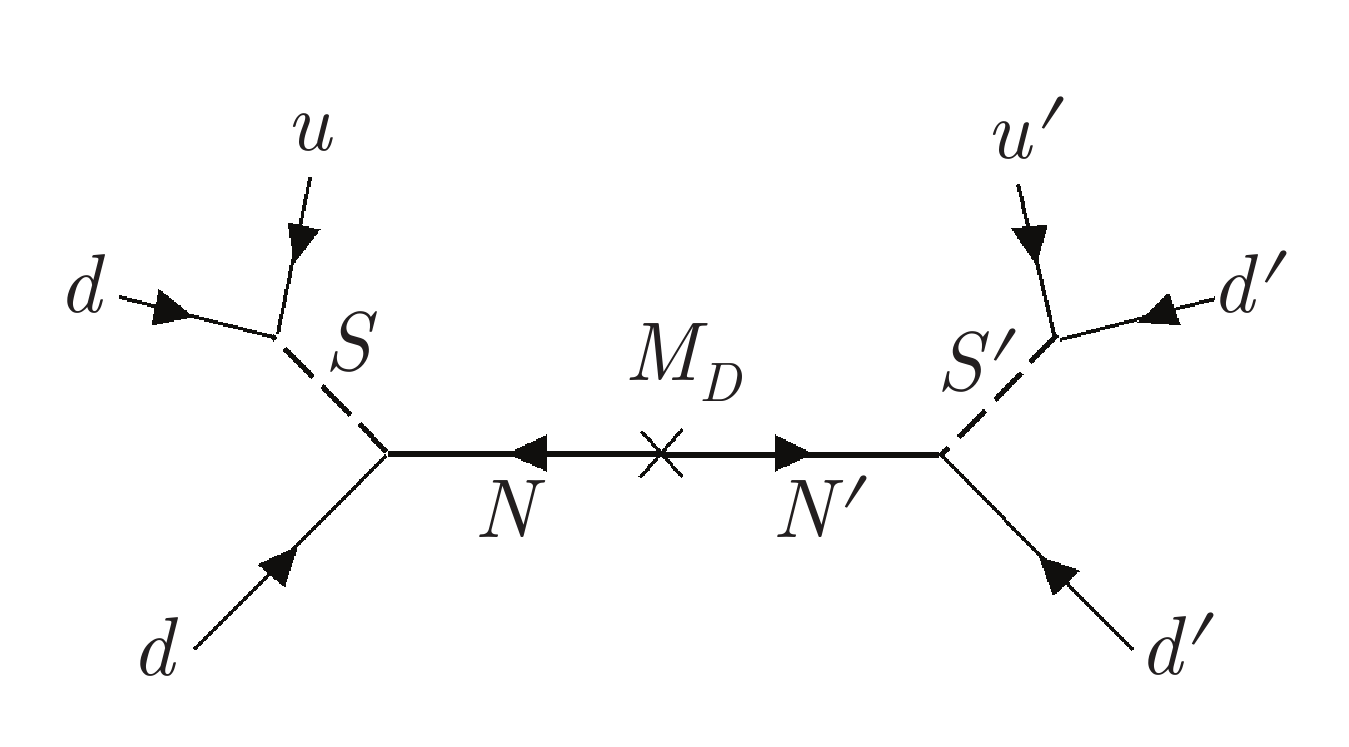}
\caption{Seesaw diagram generating $n-n'$ mixing }
\label{fig:nn'}
\end{figure}

Integrating out the heavy fermions and scalars, the diagrams shown on Fig. \ref{fig:nn'}
effectively induce operator (\ref{ops}) and 
for $n-n'$ mixing mass we get 
\be{epsilon}
\epsilon \simeq 
 \frac{g^2 h^2 N_{\rm eff} \times 10^{10}~{\rm GeV}^5}{M_S^2M_{S'}^2 M_D}  \times 10^{-10} ~{\rm MeV}
\ee
where $N_{\rm eft}$ is the effective number of $N,N'$ states which takes into account that the latter 
can have different masses, i.e. $h^2 N_{\rm eft}/M_D = \sum_a h_a^2/M^{(a)}_D$. 
The masses of $S$ and $S'$ are split due to the couplings with $Z_2$ odd scalar $\eta$, 
$\eta (S^\dagger S - S^{\prime\dagger}S')$, so that  $M_{S} \neq M_{S'}$. 
For having large enough $\epsilon$, one has to take into account the LHC limits on the 
color triplet $S$ involved in the game (for more details, see Ref. \cite{BM}). 
Namely, the first term in ${\cal L}$ (\ref{S}) induces the contact operators $\bar q q \bar q q$  
which are restricted by the compositeness limits. 
Namely, the LHC limit $\Lambda^-_{LL,RR} > 22$~TeV \cite{PDG2018} translates 
to  $(M_S/g)^2  > 0.75 \times 10^{8}~{\rm GeV}^{2}$ or so which can be saturated e.g. 
for $M_S \simeq 1.7$~TeV  and $g\simeq 0.2$. 
Therefore, $\epsilon \sim 10^{-10}$~MeV can be achieved e.g.  taking $M_{S'} \simeq 50$ GeV, 
$M_D\simeq 5$ GeV and $h^2 N_{\rm eff} \sim 10^2$. While this parameter space looks rather 
marginal, it is not excluded by the present experimental bounds.  

This mass term induces small mixing between $n-n'$, with a mixing angle $\theta = \epsilon/\Delta m$.  
For our benchmark values $\epsilon = 10^{-10}$~MeV and $\Delta m = 1$~MeV we have  $\theta=10^{-10}$. 
This mixing in turn induces transitional magnetic moment $\mu_{nn'}=\theta \mu_n$
between the mass eigenstates $n_1 = n + \theta n'$ and $n_2 = n' -\theta n$,   
Therefore, the heavier eigenstate $n_1 \approx n$ can decay into the lighter one $n_2 \approx n'$  
with the photon emission: 
\be{Gamma-photon}
\Gamma(n\to n'\gamma) = 
\frac{\vert \mu_{nn'} \vert^2 (m_n^2 - m_{n'}^2)^3}{8\pi m_n^3} = \frac{\theta^2}{\pi} \mu_n^2 \Delta m^3 
\ee 
In addition, as far as the mirror photon is also massless, $n\to n'\gamma'$ decay should take place
 with the same width, $\Gamma(n\to n'\gamma') = \Gamma(n\to n'\gamma)$   
(once again, one can neglect the difference between the ordinary and mirror magnetic moments 
and take  $\mu_n=\mu_{n'}$). Thus, the total rate of $n\to n'$ decay is
$\Gamma(n\to n') = 2\theta^2 \mu_n^2\Delta m^3/\pi = 2\mu_n^2 \epsilon^2 \Delta m/\pi$, with 
a photon $\gamma$ and mirror photon $\gamma'$ channels both having equal ratios $=1/2$.

\begin{figure}
\includegraphics[width=0.35\textwidth]{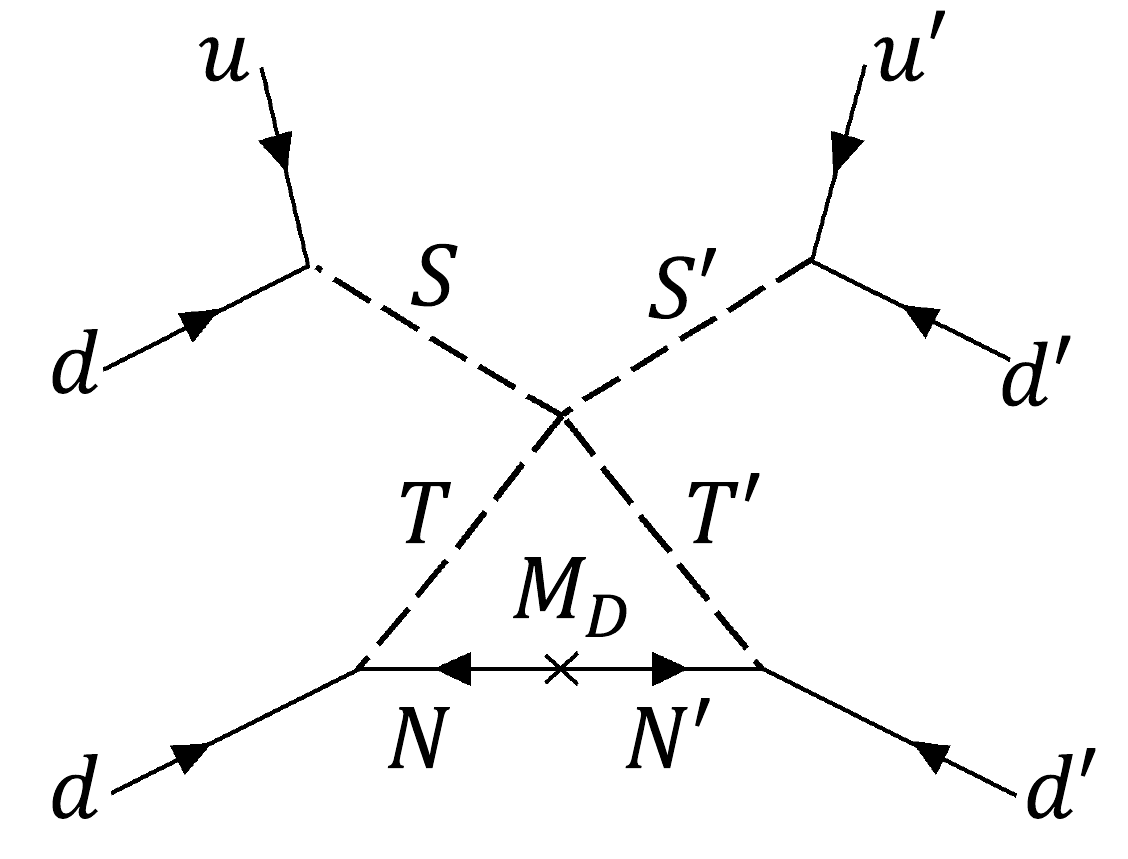}
\caption{Loop diagram generating $n-n'$ mixing through the strongly coupled $SU(N_c)$ system}
\label{fig:nn'-loop}
\end{figure}

\medskip 
\noindent 
{\bf 5.} 
There can be additional decay channels with emission of some other massless bosons. 
Let us discuss e.g. the possibility when $n-n'$ mixing emerges not at tree-level 
as in Fig. \ref{fig:nn'} but by loop mechanism shown in Fig. \ref{fig:nn'-loop}. 

Let us assume that the heavy Dirac Fermions $N$ are not gauge singlets but are multiplets 
of some gauge group $SU(N_C)$ say in fundamental representations, $N^a$ and $N'_a$, 
$a=1,2,...N_C$ being the $SU(N_C)$ index, so that we have $N_C$ Dirac fermions 
with equal masses $M_D N^a N'_a + {\rm h.c.}$. 
In this case the Yukawa terms $S^\dagger d N$ and 
$S^{\prime\dagger} d' N'$  in (\ref{S}) are forbidden by $SU(N_C)$ symmetry. 
However, one can introduce 
the additional color-triplet scalars $T^a$ and $T'_a$ 
also in fundamental representations of $SU(N_C)$,  
and modify the Lagrangian terms (\ref{S}) to the following: 
\beqn{S-T}
&& \cL = S ud + Sqq     +  T^\dagger d N  +   {\rm h.c.}   
\nonumber \\
&& \cL' =  S' u' d' + S q' q'  + T^{\prime\dagger} d' N'    +   {\rm h.c.} 
\nonumber \\
&& \cL_{\rm mix} = S^\dagger  S^{\prime \dagger} T T' + M_DN N' +   {\rm h.c.} 
\eeqn 
In this case $n-n'$ mixing is induced via the loop-diagram shown in Fig. \ref{fig:nn'-loop} 
which diagram also would induce the transitional moments 
$\kappa_{nn'}$ and $\kappa'_{nn'}$ between the neutron and mirror neutron as in  (\ref{Lagrangian}). 
One can imagine that there is also a gauge $U(1)$ symmetry in addition to $SU(N_C)$, 
with rather large coupling constant. In this way, in addition to ordinary and mirror photons, 
also a "third" photon $\gamma_3$ associated with the $U(1)$ gauge field $A^{(3)}_\mu$ enters the game. 
Then attaching the respective external photon line to the diagram of Fig. \ref{fig:nn'-loop}, one obtains 
also a transitional magnetic moment between $n$ and $n'$ related to "third" photon, 
$\frac12 \kappa_{nn'} F^{(3)}_{\mu\nu} \ov{n}\sigma^{\mu\nu} n' + $ h.c. where 
$F^{(3)}_{\mu\nu} = \partial_\mu A^{(3)}_\nu - \partial_\nu A^{(3)}_\mu$.  
In this way, there emerges an invisible decay channel $n' \to n\gamma_3$ with a width 
\be{gamma3}
\Gamma(n\to n'\gamma_3) = 
\frac{x^2}{\pi} \mu_n^2 \Delta m^3 
\ee
where $x = \kappa_{nn'}/\mu_n$ is the "third" transitional magnetic moment in units of $\mu_n$. 
For large $N_C$, large gauge constant $g_3$ of extra $U(1)$, and 
large coupling constants in (\ref{S-T}), $x$ can be comparable or even larger then $\theta$. 
In addition, the mass term induced by the loop can be suppressed by symmetry reasons 
making use of e.g. Voloshin's symmetry. In this way, the invisible decay channel $n' \to n\gamma_3$ 
can become dominant.\footnote{
Yet another invisible decay channel can be $n \to n' + \beta$ where $\beta$ is the  
Goldstone particle related to spontaneous breaking of $\rB$ and $\rB'$ baryon numbers to 
a diagonal combination $\rB+\rB'$ at some scale $V$ 
(implications of such Goldstones are discussed in Ref. \cite{BM}).  
 This massless $\beta$  interacts between $n$ and $n'$ states with the Yukawa 
coupling constant $g_{\beta} = \epsilon/V$, and thus $n\to n'\beta $ decay rate can exceed 
that of $n\to n'\gamma$ if $V < $ few GeV. 
 }

For total decay width of $n\to n'$ decay we have 
\beqn{Gamma-tot}
&&  \Gamma_{\rm new} = (1+ A_{\rm inv}) \frac{\theta^2}{\pi}  \mu_n^2 \Delta m^3 = \nonumber  \\
&&  \frac{1+ A_{\rm inv} }{6} \frac{\theta^2}{10^{-20}} \left(\frac{\Delta m}{1.57~{\rm MeV}}\right)^3 \times 
7 \cdot 10^{-30}~{\rm GeV} 
\eeqn
where the decay width is normalized to the maximal mass difference $\Delta m_{\rm max}$ (\ref{Be}) 
allowed by $^9$Be stability, and $A_{\rm inv}$ denotes effective contribution of invisible decay 
channels as $n\to n' \gamma'$, $n\to n' \gamma_3$ etc. which should be compared to the value 
(\ref{Gamma-new}) needed for explanation of the neutron lifetime discrepancy.  
The branching ratio of ``yet-invisible" decay channel (with ordinary photon $\gamma$) is
\be{Br} 
\frac{\Gamma (n\to n'\gamma)}{\Gamma_{\rm new}}  = \frac{1}{1+A_{\rm inv} } 
\ee
In particular, in the absence of "third" photon,  
and massless mirror photons coupled as ordinary one, i.e. $A_{\rm inv} =1$, 
we have ${\rm Br}(n\to n'\gamma)={\rm Br}(n\to n'\gamma') = 0.5$. 
If mirror and third photons are massive, we have  $A_{\rm inv} =0$ and   
only $n\to n'\gamma$ remains. 
(e.g. due to $Z_2$-symmetry breaking, the VEVs of two doublets $\phi'_{1,2}$ 
could break also mirror electric charge and thus render mirror photon massive). 
But for $x \gg \theta$ the invisible decay into third photon becomes large 
and the decay channel with ordinary photon becomes subdominant.

\begin{figure}
\includegraphics[width=0.5\textwidth]{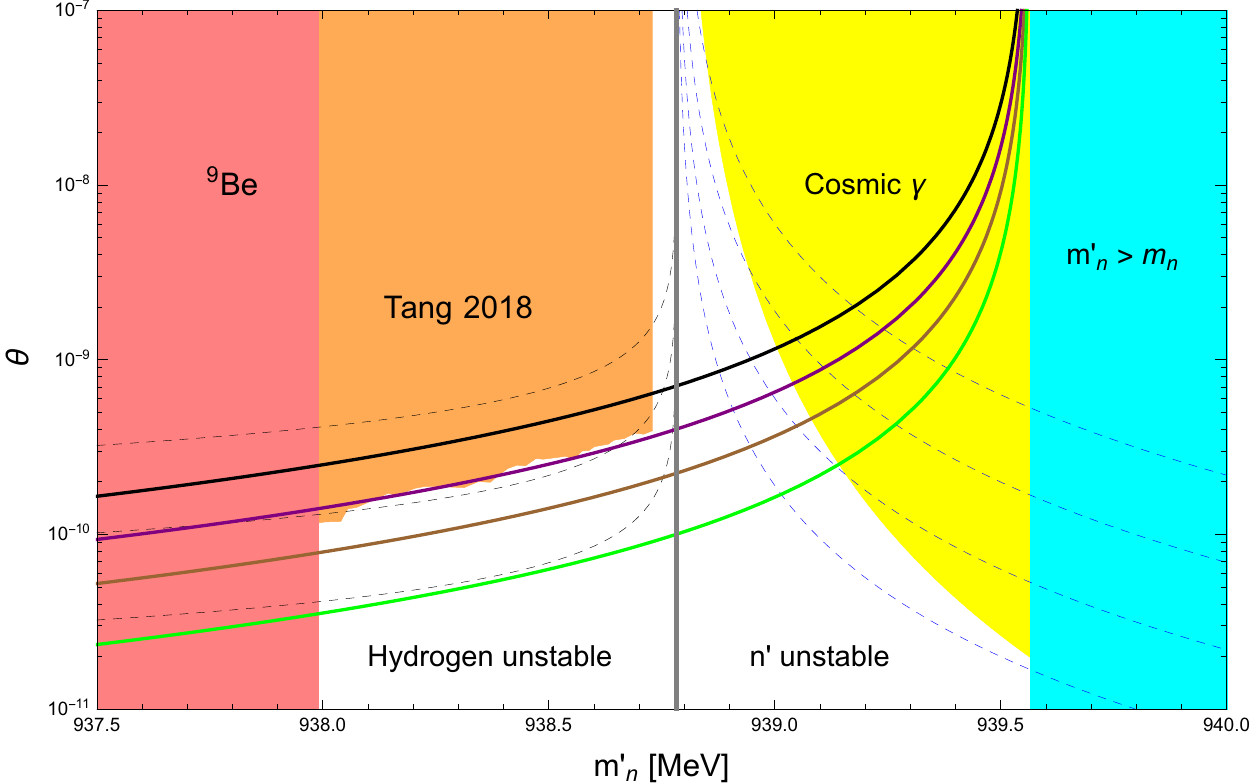}
\caption{The allowed regions for the mirror neutron mass $n'$ vs. $n-n'$ mixing angle $\theta$. 
The black, purple, brown and green solid curves, all normalised to the needed 
decay width $\Gamma_{\rm new} = 7 \times 10^{-30}$~GeV (\ref{Gamma-new}), 
correspond respectively to the $n\to n' \gamma$ branching ratios  
$\Gamma(n\to n'\gamma)/\Gamma_{\rm new} = 1, 0.5, 0.25, 0.1$. 
The vertical solid line separates regions $m_{n'} > m_p + m_e$ (unstable dark neutron $n'$) and 
$m_{n'} < m_p + m_e$ (unstable hydrogen atom). 
The black dashed curves in the latter region correspond to $^1$H  lifetimes 
 $\tau(^1{\rm H}\to n' \nu_e) = 10^{20}, 10^{21}$ and $10^{22}$ yr (from up to down), 
 while the blue dashed 
curves in the region $m'_n > m_p+m_e$ correspond to $n'$  lifetimes 
$\tau(n'\to pe \bar\nu_e) = 10^{14}, 10^{15}, 10^{16}$ and $10^{17}$ yr (again from up to down).
The shaded regions are excluded by $^9$Be stability (pink), 
by limits on cosmic $\gamma$ background in the $0.1\div1$~MeV range (yellow), 
and by photon counting in the energy range of $780\div 1600$~keV in recent experiment by Tang {\it et al.} 
(orange).  
   }
\label{fig:plot}
\end{figure}

Solid curves in Fig. \ref{fig:plot} show the parameter space (mixing angle $\theta$  vs. mirror neutron mass 
$m_{n'} = m_n - \Delta m$) needed for achieving $\Gamma_{\rm new}  = 7 \times 10^{-30}$ ~GeV 
for different $A_{\rm inv}$.  
Namely, the black solid curve corresponds to the case when $A_{\rm inv}=0$, i.e. 
only  $n\to n' \gamma$ decay is operative:  
$\Gamma (n\to n'\gamma)/\Gamma_{\rm new} = 1$.    
The solid purple corresponds to a benchmark case $A_{\rm inv}=1$ when $n\to n'$ 
decay occurs symmetrically with the ordinary and mirror photon emission, 
$\Gamma (n\to n'\gamma) = \Gamma (n\to n'\gamma')$. The brown and green curves show 
the cases when contribution of "third" photon $\gamma_3$ becomes dominant, 
respectively with $A_{\rm inv} = 3$ and $A_{\rm inv} = 9$.

\medskip 
\noindent {\bf 6.} 
Let us discuss now implications of $n - n'$ mixing and $n\to n'$ decays  
provided that $m_{n'} < m_n$ (but $m_{n'} > m_n - \Delta m_{\rm max} = 937.99$ MeV 
as it is required bu nuclear stability bound (\ref{Be})) 
which crucially depend on the mass range of dark neutron $n'$. 

Namely, if $m'_n > m_p+m_e= 938.783$~MeV, then mirror neutron $n'$ 
(more precisely, the lighter mass eigenstate $n_2 = n' - \theta n$)  
is not stable against $\beta$-decay $n' \to p+e + \bar\nu_e$. Thus, the $\beta$-decay rates 
of $n$ and $n'$ can be directly compared:  
\beqn{n'-decay} 
&& \Gamma(n \to pe\bar\nu_e)  = \frac{G_V^2  (1+3g_A^2) m_e^5}{2\pi^3} \, F\big(\frac{Q}{m_e}\big)
\nonumber \\
&& \Gamma(n' \to pe\bar\nu_e)  = \frac{\theta^2 G_V^2  (1+3g_A^2) m_e^5}{2\pi^3} \, F\big(\frac{Q'}{m_e}\big)
\eeqn
where $G_V = G_F\vert V_{ud} \vert$ is the Fermi constant corrected by Cabibbo mixing, 
 $g_A\approx 1.27$ is the axial coupling constant,  
$Q=m_n - m_p -m_e=0.7823$ MeV and $Q'= m'_n - m_p + m_e < 0.7823$~MeV  
are respective $Q$-values. 
The function 
\beqn{f}
&& F(x) =  \frac{\sqrt{x(x+2)}}{60}\big(2x^4 + 8x^3 + 3x^2 -10x -15\big) \nonumber \\
&& + \frac{x+1}{4} \ln\big(1+x+\sqrt{x^2+2x}\big)
\eeqn
describes the phase space factor for the given $Q$-value.  
%
%
Therefore, the lifetime of $n'$ can be related to the neutron lifetime 
and it can be estimated as 
\beqn{tau-n'}
&& \tau(n' \to pe\bar\nu_e) = \frac{F(Q/m_e)}{F(Q'/m_e)} \, \frac{\tau(n\to pe\bar\nu_e)}{\theta^2} = \nonumber \\
&& \frac{F(Q/m_e)}{F(Q'/m_e)} \left(\frac{10^{-10}}{\theta}\right)^2 \times 2.8 \times 10^{15}~{\rm yr}
\eeqn
The instability of dark matter is not a problem in itself once its decay time exceeds the age of the 
Universe $t_U = 1.4 \times 10^{10}$~yr. In fact, a few per cent fraction of dark 
matter decaying in invisible mode before present days could even help to reconcile the 
discrepancy between the Hubble constant value determined from the CMB measurements by Planck Satellite 
from one side, 
and its value obtained by  direct astrophysical measurements from other side \cite{BDT}.
The problem is that $n'$ decays into visible particles (proton and electron),  
together its radiative decay channel $n' \to p e\bar\nu_e \gamma$ with a branching ratio $\sim 10^{-2}$, 
would contribute to cosmic diffuse $\gamma$ background at MeV energies. 

The blue dash curves in Fig. \ref{fig:plot}  mark the parameter space which can lead to $n'$ 
decay time in the range $10^{14}-10^{17}$ yr.  Yellow shaded region corresponds to excluded 
region obtained by requiring, rather conservatively, that the $\gamma$ fluxes produced by 
these decays  should not exceed their experimental values obtained by direct observations \cite{Inoue}. 
Taking into account that main contribution to  $\gamma$-background in the MeV range 
is supposedly produced by the Seyfert galaxies and blazars, one should expect that real 
limits will be more stringent.  

Dark neutron $n'$ would be stable if its mass is small enough. 
 Namely, if $m_{n'} < m_p+m_e = 938.783$~MeV, the the decay $n' \to pe\bar\nu_e$ is 
 forbidden and $n'$ would be a stable dark particle. 
 However, this situation would imply that the hydrogen atom $^1$H (protium) should be unstable, 
 and it would decay into dark neutron and ordinary neutrino via electron capture, $p+e \to n' + \nu_e$. 
 Its decay width can be readily  estimated as\footnote{   
It should be stressed that we talk about the hydrogen atom and not its nucleus (proton). 
In fact, the proton decay $p\to n'e^+\nu_e$ would occur if $m_{n'} < m_p - m_e = 937.761$~MeV   
which is already excluded by $^9$Be stability, $m_{n'} > 937.992$~MeV (\ref{Be}).  
}
 \beqn{H-decay} 
&& \Gamma(^1{\rm H} \to n' \nu_e) = \frac{\theta^2 G_V^2  (1+3g_A^2)}{2\pi^2 a_0^3} (m_p+m_e-m'_{n})^2 
\nonumber \\ 
&& = \theta^2 \left(\frac{m_p+m_e-m_{n'}}{0.783~{\rm MeV} } \right)^2 \times 1.23 \cdot 10^{-33} ~{\rm GeV}
\eeqn
where $a_0 = (\alpha m_e)^{-1}$ is the Bohr radius. 
Thefore, for the lifetime of the hydrogen atom we get
\be{tau-H}
\tau(^1{\rm H}) = \frac{17~{\rm yr} }{\theta^2} \left(\frac{0.783~{\rm MeV} }{m_p+m_e-m_{n'}} \right)^2 
\ee
Black dashed curves in Fig. \ref{fig:plot} correspond to protium lifetimes in the range of 
$10^{20}-10^{22}$ yr. 

Surprisingly, no direct experimental limits are available on the protium dark decay 
into a dark particle $n'$ and (in practice invisible) neutrino.  
 Very existence of our universe limits the hydrogen decay time to be larger than 
 the present cosmological age $t_u = 1.4 \times 10^{10}$~yr.  
Disappearance of more than 1 \% of the hydrogen in the Universe would affect 
the Big Bang Nucleosynthesis tests.  
Thus, from the primordial hydrogen abundance 
one can infer a conservative lower bound 
$\tau(^1{\rm H} \to {\rm inv}) > 1.4 \times 10^{12}$~yr or so.  

Somewhat stronger bounds can be obtained from the electron capture process 
$p+e \to n' + \nu_e$  in the sun
which produces mono-energetic neutrinos with $E_\nu \approx m_p + m_e - m_{n'}$.  
Let us take the maximal value $E= 0.78$~MeV
which corresponds to the lower extreme $m_{n'} = 937.992$~MeV in (\ref{Be}). 
Let us assume that flux $\phi^\nu_{n'}$ of $\nu_e$ produced in this way is no larger  
than one fifth of  the  $^7$Be solar neutrino flux 
$\phi^\nu_{\rm Be} \simeq 5 \times 10^9$~cm$^{-2}$ s$^{-1}$ ($E^\nu_{\rm Be} =0.862$~MeV)
-- otherwise it would be detected by BOREXINO experiment (this assumption on the BOREXINO 
sensitivity is of course is an exaggeration).  
In other words, we infer that $\phi^\nu_{n'} < 10^9$~cm$^{-2}$ s$^{-1}$   
which puts the lower limit  on the proton lifetime $\tau_{pe}$  against the electron capture process 
$pe \to n' \nu_e$  in the sun, $\tau_{pe} > 7 \times 10^{11}$~yr or so. 
For the sake of comparison, the proton lifetime against 
the dominant reaction $pp\to de^+ \nu_e$, alias the hydrogen burning time in the sun, 
is $t_{pp}\simeq 1.2 \times 10^{10}$~yr, i.e. practically the age of the Universe. 

The rate of $p+e \to n' + \nu_e$ reaction in the sun can be obtained 
from the hydrogen atom decay rate (\ref{H-decay}) 
roughly by substituting the electron density in atom $1/a_0^3$ by the electron 
density in solar plasma $n_e = \rho X/m_p$. 
More precise calculation of the unbound electron capture rate, obtained by 
integration over the Maxwell-Boltzmann distribution of electrons, gives 
\beqn{sun}
&& \frac{\Gamma(pe\to n'\nu_e)}{\Gamma(^1{\rm H}\to n'\nu_e)}  = 
\sqrt{\frac{2\pi^3 m_e}{T}} \, \alpha \, (n_e a_0^3) \nonumber \\
&& \simeq 30 \times \left(\frac{10^6~{\rm K}}{T} \right)^{1/2}  \left(\frac{n_e}{10^{24}/{\rm cm}^{3}}\right)
\eeqn 
Taking typical temperature as $T\simeq 10^7$~K,
typical density $\rho \simeq 100$~g/cm$^3$ and the hydrogen mass fraction $X\simeq 0.5$ 
 and thus  $n_e \simeq 25\times 10^{24}$~cm$^{-3}$ in solar interior within the radius $0.2\, R_\odot$, 
we get $\Gamma(pe\to n'\nu_e)/\Gamma(^1{\rm H}\to n'\nu_e) \simeq 250$. 
Therefore, the bound $\tau_{pe} > 7 \times 10^{11}$~yr inferred for 
mono-energetic neutrinos with $E_\nu = 0.78$ MeV
is roughly equivalent to  $\tau(^1{\rm H}) > 2\times 10^{14}$~yr or so. 
Applying the same consideration for smaller values of $E_\nu = m_p + m_e - m_{n'}$, 
e.g. $E_\nu = 0.4$~MeV,  the $\nu_e$ flux produced by $p+e \to n' + \nu_e$ 
process should be compared with 
the dominant $pp$ neutrino flux, which would render this limit more than 
an order of magnitude weaker, around $10^{13}$~ yr. 

On the other hand, the reaction $p+e \to n' + \nu_e$ produces dark fermions $n'$ right 
inside the sun. Their presence, if abundant, would change the thermal conductivity of solar 
interior and this be tested by the helio-seismological data which requires a specific study. 
By inferring that the overall mass of $n'$ produced in the sun from its birth is smaller than 
e.g. $10^{-6} ~ M_\odot$, we would get $t_{pe} >  10^{15}$ yr or so which translates into 
$\tau(^1{\rm H}) > 2.5\times 10^{17}$~yr. 
Thus, it seems that the solar physics cannot exclude the protium lifetime  as large as 
$\tau(^1{\rm H}) = 10^{21}$~yr typical for our model. 

One can discuss also the electron capture processes  by nuclei 
$(Z,A)+e \to (Z-1,A-1) + n' + \nu_e$ in the electron-degenerate cores of the heavier stars. 
Let us consider e.g.  carbon white dwarfs with a central densities $\rho \sim 10^6$ g/cm$^3$  
at which densities electrons become relativistic, i.e. their 
Fermi momentum $p_F \approx 3.1 n_e^{1/3}$ 
becomes comparable to the electron mass $m_e$. However, it is not enough to overcome 
the energy threshold of $^{12}{\rm C} + e \to ^{11}{\rm B} + n' +\nu_e$ reaction which is 
above 15 MeV. The same applies to the case of oxygen, 
$^{16}{\rm O} + e \to ^{15}{\rm N} + n' +\nu_e$ which has threshold energy of about 12 MeV. 
Therefore, such processes can start only at densities approaching $10^9$ g/cm$^3$, 
as e.g. in heavy white dwarfs with masses saturating Chandrasekhar limit or 
in dense pre-supernova  cores of very massive  stars. The question whether this can have 
any observable consequences or whether it can set stronger limit on the hydrogen lifetime 
deserves special consideration.

A ``silent" disappearance  of hydrogen atom $^1{\rm H} \to n' \nu_e$ 
leaving the party without saying ``Good Bye" 
is difficult to detect experimentally. Even the daughter radiative branch $^1{\rm H} \to n' \nu_e \gamma$, 
with emission of a single photon with the energy up to 0.78 MeV, can be hardly discriminated from the 
background. The intriguing  possibility that hydrogen, the most abundant chemical element in the 
Universe constituting about 75 \% of its visible mass, can in fact be metastable remains as 
a challenge for future experiments. 

In principle,  some other elements $(Z,A)$ could also decay via the electron capture 
$ep \to n' \nu_e$. However, for any stable element (apart the hydrogen) the $^9$Be stability 
condition (\ref{Be}) does not leave an available phase space for the transition 
$(Z,A) \to (Z-1,A-1)+ n' + \nu_e$.  Perhaps it would be interesting to address exotic decays 
of some unstable proton-rich elements for which such transitions are allowed.\footnote{
Let us remark that at the lower edge of allowed range (\ref{Be}), namely for 
$m_{n'} < 938.06$~MeV,  the deuterium atom decay $^2{\rm H} \to 2n' + \nu_e + X$
becomes cinematically allowed, as a combination of electron capture $p+e \to n' + \nu_e$ 
and neutron decay $n\to n'+X$. However, due to a minuscule phase space and double 
suppression $\sim \theta^4$, the deuterium lifetime will be extremely large,  
beyond any practical interest.  For any other stable element even such a double decay is 
cinematically forbidden.  
}

Consider for example  $^{48}$Ni which has a doubly-magic nuclei $(Z=28,A=48)$, and decay time 2.1 ms.   
There is no bound nuclei $(Z=27,A=48)$ to which $^{48}$Ni could transform via $\beta^+$-decay 
or electron capture, and there is no bound nuclei $(Z=27,A=47)$ to which it can transform by 
expelling one proton. In fact, $^{48}$Ni can decay via double processes, electron capture by one 
proton accompanied  by $\beta^+$-decay of another proton, $^{48}$Ni $\to$ $^{48}$Fe $(Z=26,A=48)$
or via expelling simultaneously two protons, $^{48}$Ni $\to$ $^{46}$Fe $+2pe$ $(Z=26,A=46)$. 
The electron capture $ep \to n \nu_e$ with outflowing neutron and simultaneous expelling of proton 
 $^{48}$Ni $\to$ $^{46}$Fe $+n+p+e$ is marginally allowed by phase space but it is suppressed kinematically 
 since $m_n > m_p+m_e$. However, in the case of $m_{n'} < m_p+m_e$ such transition with 
 emission of dark neutron $n'$  will not be suppressed and thus  
  $^{48}$Ni $\to$ $^{46}$Fe transition without ordinary neutron $n$ could take place at detectable 
  level accompanied by only one proton. Unfortunately, not much is known about $^{48}$Ni decay channels. 
  Some other elements also can be of interest. E.g.  $^{50}$Co ($T_{1/2} = 38.8$~ms)  
   via electron capture $ep \to n' \nu_e$ could be transformed into $^{49}$Fe, 
  if $m_{n'} < 938.615$ MeV. Analogously $^{12}$N ($T_{1/2}=11$ ms), apart of its usual $\beta^+$ decay into 
  the stable $^{12}$C, would have a new decay channel $^{12}$N $\to$ $^{11}$C $+ n'+ \nu_e$ 
  if $m_{n'} < 938.182$ MeV, with easily detectable $^{11}$C ($T_{1/2} = 20$ m). 
  Another example, relatively stable $^{8}$B ($T_{1/2}=770$ ms) has usual $\beta^+$ decay into 
  $^{8}$Be which then promptly decays in two $\alpha$-particles. In the case $m_{n'} < 938.646$ MeV, 
  it could have a decay channel into $^7$Be.  Via the electron capture, the latter would end up in 
  $^7$Li after 53 days or so. 
  
  Concluding this section, provided that the beryllium bound (\ref{Be}) is fulfilled, 
  $n\to n'$ decay has no strong observable consequences for nuclei 
 besides the intriguing possibility that the hydrogen atom can be unstable.  
    However, it will have dramatic consequences for neutron stars (NS). 
  Given that the equation of state (EoS) of mirror nuclear matter, despite a MeV range mass 
  difference between ordinary and mirror nucleons, should be essentially the same, 
   $n\to n'$ conversion would rapidly transform the ordinary NS, after its birth, 
  into a mixed star with half of its mass constituted by mirror matter. 
  Now two components with the same EoS can be ``packed" inside the same volume 
  which changes the pressure - mass balance in the star and thus changes the 
  mass-radius relations. 
  Namely, the mixed NS will be more compact than the initial pure neutron star of the given mass. 
 with the radius of about a factor of $\sqrt2$ smaller than the initial radius of the newborn NS. 
On the other hand, also the maximal mass  of the mixed NS 
will be reduced by a factor of $\sqrt2$ with respect of the pure NS. 
For example, a realistic EoS of Ref. \cite{Sly} for pure NS can support 
$M_{\rm max} \simeq 2.1\, M_\odot$, in which case maximal mass of mixed NS 
is about $1.5\, M_\odot$; any NS with a larger mass should collapse to black hole. 
Therefore, NS of $2\, M_\odot$ could not exist unless the EoS is so stiff  
that can support  $M_{\rm max} \simeq 3 \, M_\odot$ for a pure NS. 
One can consider also a possibility that after the supernova explosion the 
newly born NS suffers a matter infall, its mass rapidly reaches a critical value and 
within days it transforms into a quark star dominantly composed of deconfined quark matter, 
which could also explain the events of delayed GRB events correlated with the supernova explosions 
\cite{Drago}. In this case, the observed  NS with larger masses reaching $2 \, M_\odot$ 
can be considered as quark stars. The implications of $n\to n'$ transition will be addressed 
in more details elsewhere \cite{Massimo}. 

\medskip 
 
\noindent  
{\bf 7}. The suggested scenario implies that the neutron 
has two decay channels, $\beta$-decay and hypothetical invisible decay.  
Therefore,  the beam experiments measure its $\beta$-decay width 
$\Gamma_\beta = \tau_{\rm beam}^{-1}$, while trap experiments  measure 
the total decay width, $\Gamma_n = \tau_{\rm trap}^{-1}$. 
However, one can question whether this hypothesis is compatible with 
other precision measurements regarding the determination of the 
Fermi constant $G_F$, the CKM mixing element $V_{ud}$ and the 
ratio of axial and vector onstants  $\lambda=g_A/g_V$. 

The neutron $\beta$-decay $n\to p e\bar\nu_e$ is described by the Fermi Lagrangian 
\be{beta-hamiltonian} 
\frac{G_V}{\sqrt2} \, \ov{p} \gamma^\mu (1 - g_A \gamma^5) n \, 
\ov{e} \gamma_\mu (1 - \gamma^5) \nu_e \, . 
\ee 
where $g_A$ is the axial coupling constant.  
In the context of the Standard Model we have $G_V=G_F \vert V_{ud} \vert$, 
where $V_{ud}$ is the CKM mixing element. 
Then, for $G_F$ determined from the muon decay, i.e. 
$G_F= G_\mu = (1.1663787 \pm 6) \times 10^{-5}$~GeV$^{-2}$ \cite{PDG2018}, 
the neutron $\beta$-decay  lifetime 
$\tau_\beta$ is given by the well-known formula 
$\tau_n \vert V_{ud} \vert^2 (1+3g_A^2) = (4908.7 \pm 1.9)~ {\rm s}$ 
which includes Coulomb corrections as well as external radiative corrections \cite{Marciano}. 
In more generic form, having in mind possible effects of new physics beyond Standard Model 
and without assuming $G_V=G_\mu \vert V_{ud}\vert $, it can be presented as 
\be{tau_n} 
\tau_\beta  = \frac{(4908.7 \pm 1.9)~ {\rm s}}{ (G_V/G_\mu)^2 (1+3g_A^2)}
\ee 
The constant $G_V$ is experimentally measured by the study of super-allowed  $0^+\to 0^+$ 
nuclear $\beta$-decays which are pure vector transitions, modulo theoretical uncertainties 
due to nuclear Coulomb effects and radiative corrections. 
For $G_F=G_\mu$ these measurements yield the world average 
$ \vert V_{ud} \vert  = 0.97417 \pm 0.00021$ \cite{PDG2016}
which value is also well-compatible with the unitarity of the CKM mixing matrix. 
In more general BSM context, 
without taking $G_V=G_\mu \vert V_{ud} \vert$, we have:  
\be{super-allowed}
(G_V/G_\mu)^2  = 0.94901\pm 0.00041
\ee
Therefore, Eq. (\ref{tau_n}) can be presented as 
\be{tau-beta} 
\tau_\beta = 
\frac{(5172.5 \pm 2.0)~ {\rm s}}{1+3g_A^2} 
\ee 
where the value (\ref{super-allowed}) is substituted for $(G_V/G_\mu)^2$. 

The axial coupling constant $g_A$ involving non-perturbative contributions 
is poorly determined theoretically.  But it is well determined  experimentally via measuring 
$\beta$-asymmetries. 
The world average $g_A= 1.2723\pm 0.0023$  reported in PDG \cite{PDG2016} then implies 
$\tau_n = 883.2 \pm 3.0$~s, compatible with both trap and beam within $1\sigma$. 
However, the average value obtained by last two most precise experiments \cite{Mund:2012,Brown:2017}
imply higher value $g_A= 1.2764\pm 0.0013$ gives $\tau_n=878.5 \pm 1.9$~s 
which is consistent with $\tau_{\rm trap}$ but it is in tension with $\tau_{\rm beam}$. 
In fact, this tension between the latest results on $g_A$ measurements \cite{Mund:2012,Brown:2017} 
and the neutron beam lifetime $\tau_{\rm beam}$ is clearly demonstrated on Fig. 6 of Ref. \cite{Brown:2017}.  
If the forthcoming experiments on $\beta$-asymmetries will confirm these results on $g_A$, 
and thus increase the tension with the value of $\tau_\beta$ determined by the beam measurements, 
then the neutron dark decay $n\to n'X$ will become useless for understanding the neutron 
lifetime anomaly. In fact, a new upper limits on the $n\to n'X$ decay rate about ten times stronger 
than (\ref{Gamma-new}), at the level $\Gamma_{\rm new} < 10^{-30}$~GeV can be firmly established, 
nearly corresponding to the  green curve in Fig. \ref{fig:plot}. 
However, the possibility of exotic neutron decay may have independent interest, 
also having in mind its  intriguing implication for  the hydrogen atom instability.  
 In fact, as one can see from Fig. \ref{fig:plot}, this limit would imply the hydrogen atom 
 lifetime of about $10^{22}$~yr.

\bigskip 

\noindent{\bf Acknowledgements} 
\vspace{2mm} 

\noindent 
This work is mainly based on my talk 
reported at the Workshop INT-17-69W 
``Neutron-Antineutron Oscillations: Appearance, Disappearance, and Baryogenesis", 
Seattle, USA, October 23 - 27, 2017  
and the slides are available at \cite{INT}.
I would like to thank the Institute of Nuclear Theory for hospitality in Seattle. 
I also thank Pierluigi Belli, Yuri Kamyshkov, Klaus Kirch, Chen-Yu Liu, David McKeen, 
Rabi Mohapatra, Valery Nesvizhevsky, Arkady Vainshtein, 
Albert Young  and other participants of the Workshop for interesting discussions and comments. 
%
Main part of this paper including  Figures was practically written between November 2017 
and  January 2018. 
I thank Riccardo Biondi and Yuri Kamyshkov for helping  in preparation of Figures. 
However, I did not rush to post this work to arXiv since in December 2017
the UCNA collaboration \cite{Brown:2017} 
reported new results on $\beta$-asymmetry measurements in the neutron decay. 
The new value of the axial  coupling constant $g_A$, determined in combination 
with  the previous PERKEO II result \cite{Mund:2012}, is inconsistent 
with the hypothesis  that the neutron lifetime discrepancy is due to the neutron invisible decay. 
Therefore, I revisited the manuscript  adding last section,  
fixed some typos and added new  references,  
and also presented this work at the NNBAR at ESS  Workshop No. 8, 
 22-23 May 2018, Institut Laue-Langevin, Grenoble \cite{ILL-May}.  

 After my Seattle talk  \cite{INT} a similar work of Fornal and Grinstein appeared \cite{Fornal}, 
with the difference that the dark particle $n'$ was considered as an elementary fermion 
with a mass chosen  {\it ad hoc}.   
It was followed by other works \cite{f1,f2,f3,f4,f5,f6,f7,f8,f9,f10,f11}   
which essentially excluded the version with elementary $n'$  
from the neutron star stability \cite{f4,f5,f6,f7}, 
and excluded $n\to n'\gamma$ decay with 1 per cent branching ratio \cite{f1}.   
In addition, Ref. \cite{f2} showed  
that the dark decay solution generically suffers from $g_A$-inconsistency problem.   
Evidently, my talk \cite{INT}  had some subconscious  impact on the community. 
It is somewhat surprising that non of these works  mentioned about it  
though some authors of Refs. \cite{f1,f4}  were present at my talk in Seattle 
and some others were informed about it.  

I was uncertain about publishing this work,   mainly because of understanding that  
the neutron dark decay solution is incompatible with 
the value of axial coupling $g_A$.  In addition, for  a while   
I found  $g_A$ consistent  solution to the neutron lifetime puzzle based on $n-n'$ oscillation  \cite{nn'}. 
Finally  I decided that the present  work can be anyway published, at least for 
the intriguing possibility of the hydrogen metastability which was  overlooked 
in previous papers. Perhaps  the exotic  decays  of the neutron and hydrogen 
may work in some future (one never knows) towards understanding 
the fundamental physics behind the neutron.



\end{document}